\documentclass[manuscript, screen]{acmart}
\AtBeginDocument{%
  }

\usepackage{graphicx} % Required for inserting images
\usepackage{xspace}
\usepackage{tabularray}
\usepackage{tcolorbox}
\usepackage{multirow}
\usepackage{booktabs}
\usepackage{pgfplots}
\usepackage{pgfplotstable}
\usepackage{adjustbox}
\usepackage{subcaption}
\usepackage{adjustbox}
\usepackage{colortbl}
\usepackage{listings}
\usepackage{algorithm}
\usepackage{algorithmic}
\usepackage{amsfonts}
\usepackage{longtable} % For tables spanning multiple pages if needed
\usepackage{comment}
\usepackage{float}
\usepackage{hyperref}

\usepackage{soul}
\usepackage{hyperref}
\usepackage{multirow}
\usepackage{listings}
\usepackage{fancybox}
\usepackage{enumitem}
\usepackage{graphicx}
\usepackage{color}
\usepackage{array}
\usepackage{amsmath}
\usepackage{xspace}
\usepackage{url}
\usepackage{tikz}
\usepackage{caption}
\usepackage{pgfplots}
\usepackage{balance}
\usepackage{subcaption}
\pgfplotsset{compat=1.16}
\usepackage{pgf-pie}
\usetikzlibrary{pgfplots.statistics,calc}
\usepackage{algorithm}
\usepackage{algorithmic}
\usepackage{adjustbox}
\usepackage{makecell}

\usepackage{tikz}
\usepackage{calc}
\def\checkmark{\tikz\fill[scale=0.4](0,.35) -- (.25,0) -- (1,.7) -- (.25,.15) -- cycle;}

\usepackage{tcolorbox}
\makeatletter
\newcommand{\mybox}[1]{%
	\setbox0=\hbox{#1}%
	\setlength{\@tempdima}{\dimexpr\wd0+13pt}%
	\begin{tcolorbox}[boxrule=0.5pt, colback=white, arc=4pt,
		left=6pt,right=6pt,top=6pt,bottom=6pt,boxsep=0pt]
		#1
	\end{tcolorbox}
}

\definecolor{songcolor}{RGB}{191,191,191}
\newcommand{\todoc}[2]{{\textcolor{#1}{\textbf{#2}}}}
\newcommand{\todogreen}[1]{\todoc{green}{\textbf{[[#1]]}}}

\newcommand{\hung}[1]{\todogreen{Hung: #1}}

\title{RGFL: Reasoning Guided Fault Localization for Automated Program Repair Using Large Language Models}

\author{Melika Sepidband}
\email{melikasp@yorku.ca}
\affiliation{%
  \institution{York University}
  \city{Toronto}
  \country{Canada}
}
\author{Hamed Taherkhani}
\email{hamedth@yorku.ca}
\affiliation{%
  \institution{York University}
  \city{Toronto}
  \country{Canada}
}
\author{Hung Viet Pham}
\email{hvpham@yorku.ca}
\affiliation{%
  \institution{York University}
  \city{Toronto}
  \country{Canada}
}
\author{Hadi Hemmati}
\email{hemmati@yorku.ca}
\affiliation{%
  \institution{York University}
  \city{Toronto}
  \country{Canada}
}

\begin{document}

\begin{abstract}
    Fault Localization (FL) is a critical step in Automated Program Repair (APR), and its importance has increased with the rise of Large Language Model (LLM)-based repair agents. In realistic project-level repair scenarios, software repositories often span millions of tokens, far exceeding current LLM context limits. Consequently, models must first identify a small, relevant subset of code, making accurate FL essential for effective repair.
We present a novel project-level FL approach that improves both file- and element-level localization. Our method introduces a hierarchical reasoning module that (i) generates structured, bug-specific explanations for candidate files and elements, and (ii) leverages these explanations in a two-stage ranking scheme combining LLM-based and embedding-based signals. We further propose a counterfactual upper-bound analysis to quantify the contribution of each localization stage to repair success.
We evaluate our approach on Python and Java projects from SWE-bench Verified, Lite, and Java. Compared to state-of-the-art baselines, including Agentless and OpenHands, our method consistently improves localization accuracy. On SWE-bench Verified, file-level Hit@1 improves from 71.4\% to 85\%, and MRR from 81.8\% to 88.8\%. At the element level, Exact Match under top-3 files increases from 36\% to 69\%. Integrating our localization into Agentless yields a 12.8\% end-to-end repair success improvement.
\end{abstract}

%\begin{IEEEkeywords}

%\end{IEEEkeywords}



\maketitle

\section{Introduction}

With the rise of Large Language Models (LLMs), new opportunities have emerged for leveraging their contextual understanding in software engineering tasks, including Automated Program Repair (APR).\cite{chen2021evaluating, wang2021codet5, fan2023large} 
Nowadays LLM-based APRs\cite{xia2023automated, xia2022less} can resolve real issues reported in GitHub repositories of well-known products and libraries, where a fix requires patching multiple files across the repository. One of the major challenges in this type of complex APR task is to find the relevant code for the LLM to focus on, since providing the entire codebase to the model is impossible. For instance, the largest repository in the well-known SWE-bench Verified\cite{jimenez2023swe} APR benchmark is nearly 12M tokens, in the SymPy project. These sizes far exceed the context limits of current LLMs. Even in cases where a repository could technically fit into the context window, presenting the entire codebase is usually inefficient and often counterproductive, since the model becomes overwhelmed with irrelevant details, making it difficult to focus on the root cause of the bug. This perfectly motivates the necessity of an accurate Fault Localization(FL), not only as a way to reduce the search space for APR, but also as a practical mechanism to adapt LLMs to large, real-world software systems. By directing the model’s attention to the most suspicious files and elements, localization ensures that the limited context budget is spent on the most relevant information.

In general, FL is a critical step in APR, which guides the repair system to the parts of the codebase most likely to contain the root cause of a bug.\cite{monperrus2018living, wong2016survey} However, traditional FL techniques, including static analysis\cite{zheng2006value}, classic information retrieval (IR)\cite{zhou2012should, saha2013improving}, and spectrum-based methods\cite{jones2005empirical}, often fail to capture deeper semantic links between the bug report and the code. These techniques typically rely on shallow heuristics, which may overlook the actual cause when the bug stems from subtle logic errors or domain-specific behavior.

Some recent work has explored the use of LLMs for FL by prompting them to identify relevant parts of the code based on their similarity to the bug report\cite{xia2024agentless} or by generating predictions directly from code context\cite{meng2024empirical}. However, these approaches typically rely on textual matching or LLM-based retrieval without explicitly asking the model to reason about why a particular code is relevant. As a result, their potential for supporting semantic or logically relevant code to an issue is limited.

\begin{figure*}[t]
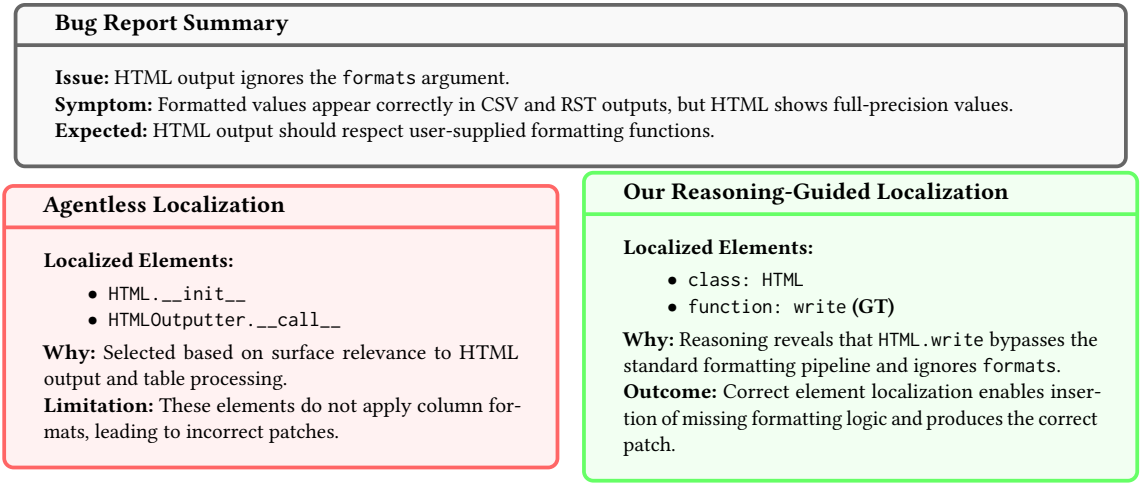

\centering
\begin{minipage}{0.98\linewidth}
\begin{tcolorbox}[
  colback=gray!5,
  colframe=black!60,
  title=Bug Report Summary,
  colbacktitle=gray!5, coltitle=black,
  fonttitle=\bfseries
]
\small
\textbf{Issue:} HTML output ignores the \texttt{formats} argument.  

\textbf{Symptom:}  
Formatted values appear correctly in CSV and RST outputs, but HTML shows full-precision values.

\textbf{Expected:}  
HTML output should respect user-supplied formatting functions.
\end{tcolorbox}
\end{minipage}
\hfill
\begin{minipage}{0.49\linewidth}
\begin{tcolorbox}[
  colback=red!5,
  colframe=red!60,
  title=Agentless Localization,
  colbacktitle=red!5, coltitle=black,
  fonttitle=\bfseries
]
\small
\textbf{Localized Elements:}
\begin{itemize}
  \item \texttt{HTML.\_\_init\_\_}
  \item \texttt{HTMLOutputter.\_\_call\_\_}
\end{itemize}

\textbf{Why:}  
Selected based on surface relevance to HTML output and table processing.

\textbf{Limitation:}  
These elements do not apply column formats, leading to incorrect patches.
\end{tcolorbox}
\end{minipage}
\hfill
\begin{minipage}{0.49\linewidth}
\begin{tcolorbox}[
  colback=green!5,
  colframe=green!60,
  title=Our Reasoning-Guided Localization,
  colbacktitle=green!5, coltitle=black,
  fonttitle=\bfseries
]
\small
\textbf{Localized Elements:}
\begin{itemize}
  \item \texttt{class: HTML}
  \item \texttt{function: write} \textbf{(GT)}
\end{itemize}

\textbf{Why:}  
Reasoning reveals that \texttt{HTML.write} bypasses the standard formatting pipeline and ignores \texttt{formats}.

\textbf{Outcome:}  
Correct element localization enables insertion of missing formatting logic and produces the correct patch.
\end{tcolorbox}
\end{minipage}

\caption{Motivating example from SWE-bench Verified (Astropy library) showing that reasoning-guided element localization identifies the true fault-inducing method (\texttt{HTML.write}), while Agentless selects only superficially related elements.}
\label{fig:motivation-example}
\end{figure*}

Consider figure~\ref{fig:motivation-example}, which illustrates a real bug report from the SWE-bench dataset~\cite{jimenez2023swe} which is from Astropy’s HTML table writer. The bug report states that when exporting a table to HTML, the user-supplied \texttt{formats} argument is ignored: values appear with full precision in HTML output, while the same \texttt{formats} argument works correctly for other writers such as CSV and reStructuredText. The ground-truth patch confirms that the fault lies in the HTML writing path: the fix adds two missing initialization steps inside \texttt{HTML.write}, namely setting \texttt{self.data.cols = cols} and invoking \texttt{self.data.\_set\_col\_formats()}, ensuring that column formatting is properly configured before generating the HTML cells.

This issue is challenging for element localization because the bug report does not explicitly name the faulty method; it describes a semantic discrepancy between formats that ``respect'' \texttt{formats} and HTML that does not. In Agentless\cite{xia2024agentless}, element localization is performed by directly asking an LLM to list functions related to the bug report, without requiring an explicit rationale tied to code behavior. In this example, Agentless selects \texttt{HTML.\_\_init\_\_} and \texttt{HTMLOutputter.\_\_call\_\_}. While these elements appear contextually relevant (initializing HTML options and processing column structures), they are not the causal point where \texttt{formats} should be applied. As a result, the localized context emphasizes superficially related components, making it harder for the repair model to identify that the missing formatting logic must be introduced in the HTML writer itself.

In contrast, we propose a reasoning-guided approach that first elicits element-level reasoning (what each candidate element does and how it relates to the observed symptom) and then uses this reasoning to guide element selection. For the same bug, our localized elements include \texttt{class: HTML} and, crucially, \texttt{function: write}, which matches the ground-truth faulty element. The generated reasoning explicitly links the symptom to the implementation: \texttt{HTML.write} constructs the HTML table body by iterating over \texttt{col.info.iter\_str\_vals()}, a pathway that formats values using the column’s default settings and bypasses the writer’s centralized formatting mechanism. This reasoning makes clear that the writer never applies the \texttt{formats} dictionary provided to \texttt{Table.write}, explaining why HTML output ignores formatting while other writers behave as expected.

This example illustrates why reasoning is valuable for localization: it shifts the model from selecting elements that merely sound related to selecting elements that are responsible for the failure. By making the causal chain explicit, our approach reliably surfaces \texttt{HTML.write} as the fault-inducing element, enabling the repair model to add the exact missing initialization steps and generate the correct patch. In contrast, without reasoning, the localization stage may over-prioritize nearby but non-causal elements, leading to incorrect edits and failed repairs even when the bug report is clear at a high level.

Building on this example, we generalize the idea into a reasoning-guided fault localization (RGFL) approach for APR pipelines. RGFL is based on the observation that LLM-as-Judge ranking over many files or code elements is effective for filtering candidates, but often insufficient for precisely identifying the most relevant ones. To address this, RGFL introduces an explicit per-candidate reasoning stage: for each file or element individually, the LLM explains its functionality, and this explanation is causally related to the reported symptom. These reasoning outputs are then used as ranking signals. The core intuition is that, by relying on natural language reasoning rather than raw code similarity, the LLM can better capture the semantic and causal connection between the bug report and potential fault locations, leading to more accurate localization.

We integrate our reasoning-based localization strategy into Agentless\cite{xia2024agentless}, a state-of-the-art open-source modular APR framework, which employs LLMs throughout the pipeline and relies on an LLM-as-Judge paradigm to rank multiple candidates at once based on their perceived relevance to the bug report, following an initial retrieval step based on embedding similarity. Our method replaces the LLM and embedding-based file and element ranking components with reasoning-informed ranking, while keeping the line localization and patch generation steps unchanged. This design allows us to isolate the impact of reasoning on the quality of localization and final repair success. 

\subsection{Research Questions}

We evaluate our approach through the following research questions:

\begin{itemize}
    \item RQ1: Can LLM-generated reasoning about source code files and elements improve file localization and element localization performance in a modular APR system like Agentless?
    \item RQ2: How generalizable is our fault localization approach across different datasets and programming languages, and how does it compare to existing baseline methods?
    \item RQ3: Does ranking files and elements based on LLM reasoning improve the success rate of final program repair, compared to SOTA APR agents?
\end{itemize}

\subsection{Contributions}

Our contributions are as follows:

\begin{enumerate}
    \item We design a novel hierarchical reasoning-guided module for fault localization, where the LLM generates explanations for the relevance of code files and elements to a bug report and ranks them accordingly. We also use an embedding-based ranking for file localization to compare with the LLM-based ranking.
    \item We integrate this reasoning module into the modular APR system, Agentless, and systematically evaluate its effect on file and element localization. We provide a systematic ablation and counterfactual study that isolates file vs element and quantifies their individual contribution to end-to-end repair, including upper bounds given perfect localization.
    \item We demonstrate that RGFL improves localization accuracy on multiple benchmarks: SWE-bench Verified, SWE-bench Lite, and SWE-bench Java\cite{zan2024swe, zan2025multi} for both Python and Java projects.
    \item We conduct experiments with multiple reasoning-capable LLMs—including Gemini 2.5 Pro\cite{comanici2025gemini}, Claude 4 Sonnet\cite{claude4sonnet}, and OpenAI’s o4-mini\cite{o3o4mini}—and evaluate the repair phase using Gemini 2.5 Pro, enabling direct comparison with SOTA agents on the SWE-bench leaderboard.
    \item We introduce a fine-grained error taxonomy for unresolved instances, distinguishing failures at the file, element, line, and repair stages. Using controlled counterfactual injections of ground-truth information at each level, we estimate upper-bound performance and isolate how much of the residual gap is attributable to fault localization versus patch generation.
    
\end{enumerate}
\section{Background and Related Works}

\subsection{Automated Program Repair (APR)}

APR aims to automatically generate patches that resolve software bugs, typically using a pipeline composed of three major stages: Fault Localization(FL), Patch Generation, and Patch Validation.\cite{le2019automated, zhang2023survey} The first step—FL—is especially critical, as it narrows the search space by identifying potentially buggy code regions that the repair system should focus on.\cite{assiri2017fault, pearson2017evaluating} If this step is inaccurate, even powerful patch generation models are unlikely to succeed.

APR systems can be broadly categorized into pipeline and end-to-end approaches. pipeline agents, such as Agentless, explicitly separate the fault localization and patch generation stages. These systems first identify suspicious code regions using dedicated techniques, and then generate candidate patches for the localized elements. This modularity allows for targeted improvements to individual components, such as localization. In contrast, end-to-end agents like OpenHands\cite{wang2024openhands} treat the repair task as a monolithic problem, prompting an LLM to directly generate patches based on the bug report and codebase context without explicit localization. While end-to-end approaches benefit from simplicity and tight integration, they may struggle in complex codebases where localization is crucial for narrowing down the search space. Our work builds on the pipeline design by improving the localization stage through LLM-based reasoning, aiming to enhance the overall repair performance.

\subsection{Fault Localization Techniques and the Role of LLMs}

FL has traditionally relied on techniques such as Spectrum-Based Fault Localization (SBFL), which uses test coverage and execution information to assign suspicion scores to program elements\cite{abreu2007accuracy, abreu2009practical, sohn2017fluccs}, and Information Retrieval (IR)-based methods that compute lexical or semantic similarity between bug reports and code (e.g., using TF-IDF, BM25, or static embeddings).\cite{zhou2012should, saha2013improving, wang2015evaluating} More recently, learning-based approaches have emerged, including deep learning and graph neural networks (e.g., DeepFL\cite{pearson2017evaluating}, LOCAGENT\cite{chen2025locagent}), which learn representations of code structure and execution behavior to improve localization accuracy.

Despite these advances, traditional methods often fail to capture the rich semantic relationships between bug reports and code. They rely on surface-level similarity or rigid structural patterns and are not easily adaptable to ambiguous bug reports that require deeper reasoning. LLMs offer a promising new direction.\cite{fan2023automated, kang2024quantitative, qin2024agentfl}

Retrieval-Augmented Generation (RAG)\cite{lewis2020retrieval} is a technique that enhances LLMs by combining them with an external retrieval system. Instead of relying only on the knowledge encoded in the model’s parameters, RAG retrieves relevant documents, code snippets, or knowledge base entries from an external source and feeds them into the LLM as additional context. This hybrid approach allows the model to ground its responses in up-to-date, task-specific, or domain-specific information.

\subsection{LLM-based APR Agents}

Recent surveys on agentic software issue resolution highlight that LLM-based approaches to APR vary along multiple orthogonal dimensions, including scaffold design (end-to-end vs.\ pipeline-based), the role of fault localization, and the use of RAG for repository-scale reasoning \cite{bouzenia2024repairagent, jiang2025agentic}. Following this taxonomy, we organize prior work into three categories most relevant to our setting.

\paragraph{Agent-based end-to-end APR systems:}
End-to-end agent-based systems treat program repair as a unified task, allowing an LLM to iteratively explore the repository, reason about the bug, and generate patches. \textbf{SWE-Agent} \cite{yang2024swe} equips LLMs with a structured agent-computer interface for navigation, editing, and execution, enabling more effective repository interaction. \textbf{OpenHands} \cite{wang2024openhands} similarly provides rich tooling and execution capabilities, prompting the LLM with large portions of the codebase and issue description to directly propose fixes. \textbf{TRAE} \cite{gao2025trae} further emphasizes aggressive tool-based exploration, allowing the model to inspect files, search symbols, and run tests before producing patches. While effective in some cases, these approaches often rely on implicit localization and may struggle to scale when precise fault isolation is required.

\paragraph{Pipeline-based APR systems:}
Pipeline-based approaches explicitly separate fault localization from patch generation, enabling targeted improvements to individual stages. \textbf{Agentless} \cite{xia2024agentless} exemplifies this design with a lightweight three-phase pipeline (localization, repair, validation) and hierarchical localization (file $\rightarrow$ function $\rightarrow$ edit location) using a combination of IR retrieval and LLM prompting. \textbf{AutoCodeRover} \cite{zhang2024autocoderover} follows a two-stage pipeline in which the LLM navigates the repository via program-structure-aware search APIs before generating candidate patches; when tests are available, it incorporates spectrum-based fault localization to refine localization. These systems demonstrate that strong localization is critical for effective downstream repair.

\paragraph{LLM-based fault localization and issue localization methods:}
Beyond full APR pipelines, a growing body of work focuses specifically on the fault localization stage. Recent surveys identify two broad paradigms for LLM-based localization: fine-tuning–based approaches and RAG–based approaches. In this work, we focus on the latter. Within RAG-based localization, prior methods can be further categorized by their retrieval strategy into \emph{vector-based RAG}, \emph{graph-based RAG}, and \emph{navigation-based RAG} approaches \cite{jiang2025agentic}.

Vector-based methods retrieve code elements using embedding similarity between bug reports and code (e.g., \textbf{BLAZE}\cite{chakraborty2025blaze}, \textbf{CoSIL}\cite{jiang2025cosil}). Graph-based approaches, such as \textbf{RepoGraph} \cite{ouyang2024repograph} and \textbf{LocAgent} \cite{chen2025locagent}, construct repository-level graphs (e.g., call or dependency graphs) and retrieve relevant context via graph traversal. Navigation-based methods allow an LLM agent to actively explore the repository to gather relevant context; \textbf{OrcaLoca} \cite{yu2025orcaloca} is a representative example, introducing priority-based scheduling, relevance scoring, and distance-aware context pruning to improve function-level localization accuracy. Hybrid approaches, such as \textbf{KGCompass} \cite{yang2025enhancing}, combine semantic graphs with retrieval to enable multi-hop localization.

However, existing RAG-based localization methods typically score or retrieve candidates based on 
similarity, graph proximity, or agent-driven exploration, without introducing an explicit reasoning step 
that evaluates each candidate independently in isolation.

\paragraph{Positioning of our approach:}
Table~\ref{tab:taxonomy_comparison} positions our method with respect to prior LLM-based APR and fault localization approaches along six key design dimensions.
\emph{Scaffold} distinguishes end-to-end agentic systems, which tightly couple localization and repair, from pipeline-based approaches that explicitly separate fault localization from patch generation, as well as FL-only methods that focus solely on localization.
\emph{Explicit FL} indicates whether fault localization is treated as a first-class stage rather than being implicitly handled during generation.
\emph{RAG Type} captures how candidate code elements are obtained, including vector-based retrieval, graph-based traversal, navigation-based exploration, hybrid combinations, and Reasoning guided.
\emph{Repo Navigation} reflects whether a method actively explores the repository using agentic tools.
\emph{Hierarchical Localization} denotes whether localization is performed across multiple granularities, such as file-, function-, and line-level.
Finally, \emph{Explicit Reasoning} indicates whether the method explicitly reasons about candidate code elements during localization, rather than relying solely on implicit scoring or similarity signals.

While some pipeline-based approaches employ hierarchical localization, none introduce an explicit per-candidate reasoning step that evaluates each candidate \emph{independently and in isolation} prior to ranking.
Our approach fills this gap by incorporating a \emph{reasoning-first, selective localization strategy} within a pipeline-based APR framework: given a fixed set of candidate files or code elements, the LLM first generates structured rationales for each candidate individually, and these reasoning outputs are then used as the ranking signal. By incorporating this mechanism into Agentless, we isolate the effect of reasoning-guided localization while remaining compatible with existing repair stages.

\begin{table}[t]
\centering
\caption{Taxonomy and feature comparison of LLM-based APR and fault localization methods. Prior approaches differ in scaffold design and retrieval strategy, while our method uniquely introduces explicit per-candidate LLM reasoning as a localization signal within a modular pipeline.}
\scriptsize
\setlength{\tabcolsep}{2.5pt}
\begin{tabular}{lcccccc}
\toprule
\textbf{Method} &
\textbf{Scaffold} &
\textbf{Explicit FL} &
\textbf{RAG Type} &
\textbf{Repo Navigation} &
\textbf{Hierarchical Loc.} &
\textbf{Explicit Reasoning} \\
\midrule
SWE-Agent \cite{yang2024swe}        & End-to-end & -- & Navigation & \checkmark & -- & -- \\
OpenHands \cite{wang2024openhands} & End-to-end & -- & Navigation & \checkmark & -- & -- \\
TRAE \cite{gao2025trae}             & End-to-end & -- & Navigation & \checkmark & -- & -- \\
\midrule
AutoCodeRover \cite{zhang2024autocoderover} & Pipeline & \checkmark & Navigation & \checkmark & -- & -- \\
Agentless \cite{xia2024agentless} & Pipeline & \checkmark & Hybrid & \checkmark & \checkmark & -- \\
\midrule
RepoGraph \cite{ouyang2024repograph} & FL-only & \checkmark & Graph & -- & -- & -- \\
LocAgent \cite{chen2025locagent}     & FL-only & \checkmark & Graph & -- & \checkmark & -- \\
OrcaLoca \cite{yu2025orcaloca}       & FL-only & \checkmark & Navigation & \checkmark & -- & -- \\
KGCompass \cite{yang2025enhancing}   & FL-only & \checkmark & Hybrid & \checkmark & -- & -- \\
\midrule
\textbf{Ours} & Pipeline & \checkmark & Reasoning-guided
 & \checkmark & \checkmark & \textbf{\checkmark} \\
\bottomrule
\end{tabular}
\label{tab:taxonomy_comparison}
\end{table}
\section{Methodology}

This work investigates the role of LLM-generated reasoning in enhancing fault localization, a critical step in the broader pipeline of APR. While prior work has demonstrated the utility of retrieval- or embedding-based approaches, the potential of explicit reasoning—where LLMs explain why specific code regions are relevant to a bug—remains underexplored. Our methodology is designed to systematically evaluate whether incorporating reasoning can improve both localization accuracy and downstream repair performance in modular APR systems.

We build upon Agentless, a state-of-the-art modular APR framework that ranks among the top-performing systems on the SWE-bench leaderboard. Agentless follows a multi-stage repair pipeline:

\begin{enumerate}
    \item File-level localization: retrieving source files likely to contain the bug;
    \item Element-level localization: identifying relevant classes, functions, and global variables within candidate files;
    \item Line-level localization: narrowing down to the specific faulty lines of code;
    \item Repair and selection: generating candidate patches and validating them against the test suite.
\end{enumerate}

Each stage leverages LLMs for information extraction, retrieval, and generation. Agentless's modularity allows us to isolate and modify specific components without affecting the overall pipeline—making it ideal for studying the impact of LLM reasoning on localization performance. However, our method does not solely rely on Agentless and can be integrated with any hierarchical FL system (OpenHands, AutoCodeRover, LocAgent, etc.). It requires only a list of candidate files, on top of which our reranking for file and element localization can be applied.

\subsection{Reasoning-based Localization Strategy}
\begin{figure*}
    \centering
    \includegraphics[width=1.0\linewidth]{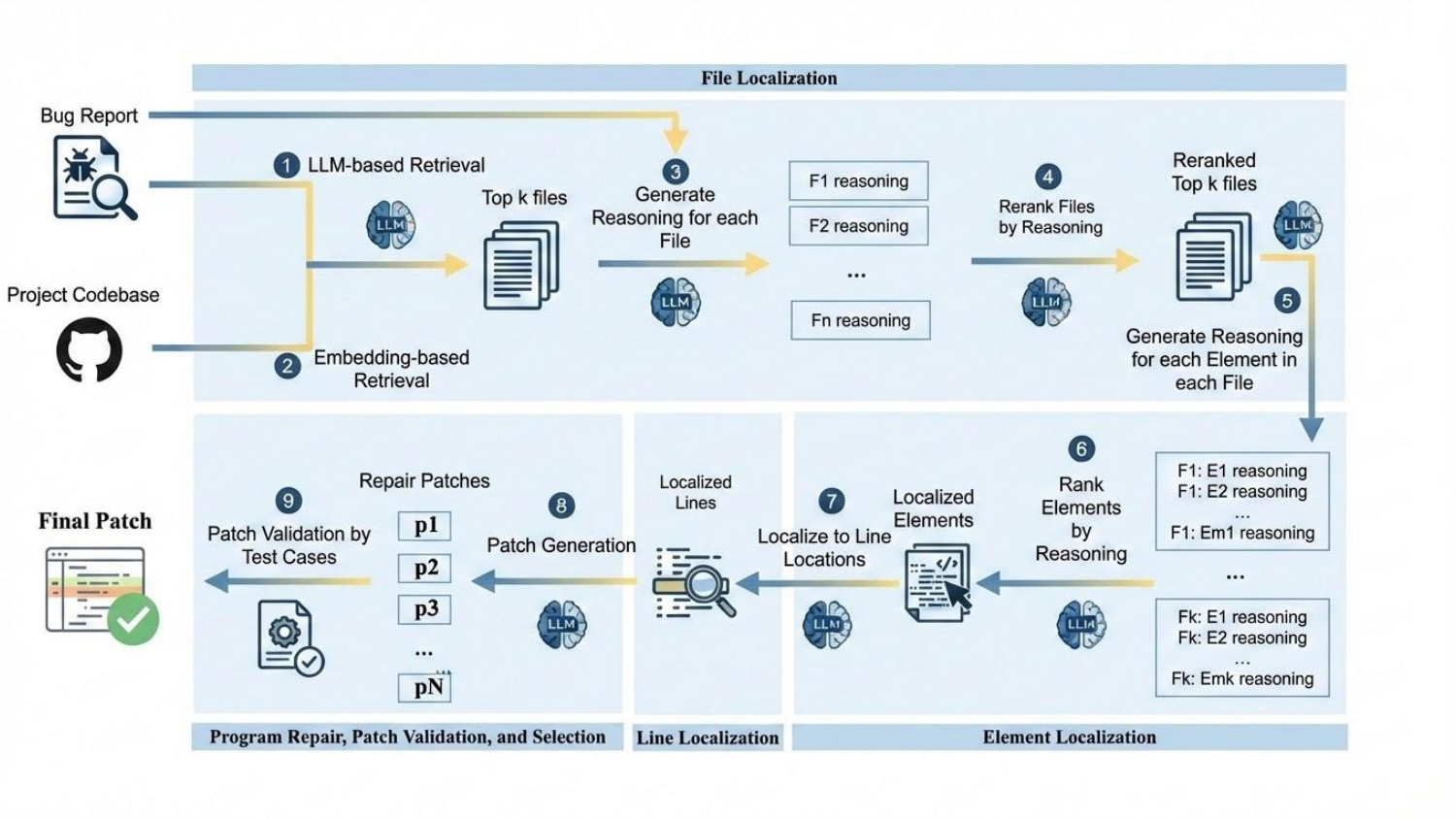}
    \caption{Overview of our reasoning-based fault localization in APR
    }
    \label{fig:fig1}
\end{figure*}

We propose a reasoning-augmented alternative to the file and element localization stages. Instead of retrieving files or elements solely based on lexical or embedding similarity to the bug report, we prompt an LLM to generate natural language reasoning about the relevance of code files or elements to the bug report. The files or elements are then ranked based on this generated reasoning. To generate reasoning, we present the LLM with one file or code element at a time, together with the problem statement. Rather than providing the entire repository and asking the model to select relevant components, we prompt it to explain the behavior of a single candidate in isolation. This design prevents the LLM from being overwhelmed by large code contexts and encourages focused, fine-grained reasoning about each file or element. In contrast, many prior LLM-based agents supply a large set of files or elements simultaneously and ask the model to assess relevance, which can dilute attention and bias the model toward surface-level similarity. Figure~\ref{fig:fig1} shows the overview of our method. Below is the detailed explanation of the methodology.\\

\noindent\textbf{File-level Localization:} We first follow the file-level localization strategy in Agentless. Generating reasoning across every file in a repository (often exceeding 1,000 files) would be computationally very expensive. By restricting reasoning to the top-k candidates identified by Agentless, we balance efficiency with effectiveness, ensuring that reasoning enhances retrieval without incurring prohibitive overhead. Agentless starts by using the issue description and the project’s codebase as input. First, it turns the codebase into a tree-like structure that shows how the files are organized in the project. Then, using this structure and the issue description, it asks the LLM to find and rank the top k files that are most likely related to the issue (step 1 in Figure~\ref{fig:fig1}). To complement prompting-based localization (which relies only on file names), it also retrieves files that contain code similar to the issue description using an embedding-based search. To keep it efficient, it filters out irrelevant folders first, then embeds the remaining code chunks and the issue description to retrieve files whose code is most semantically similar to the bug report (2). Finally, it combines the files found by the LLM and the retrieved files to create the final list of suspicious files. Then we generate reasoning for the files retrieved by Agentless (rather than all files in the repository, which would be prohibitively expensive for large repositories).(3)
Specifically, we prompt the model to first reason about the functionality of each file, and then rank them according to their relevance to the given bug report. (4) 

\noindent\textit{File-Level Reasoning Prompt:}
We begin by asking the LLM to generate a natural language explanation of a file’s functionality in the context of the bug report. This is used later for semantic ranking:
\begin{tcolorbox}[colback=gray!5!white, colframe=black!75!white, boxrule=0.5pt, arc=2pt, fontupper=\ttfamily\footnotesize, left=2pt, right=2pt, top=2pt, bottom=2pt, ]
A user is trying to fix a bug described in the following report:
\{bug\_report\}

Below is a code file from a repository:
\{file\_content\}

Explain the purpose and functionality of this code in the context of the bug report.
Focus on what this file does and whether it may be related to the bug.
\end{tcolorbox}
\noindent\textit{File Ranking Prompt:}
Once reasoning is generated for each file, we prompt the LLM to rank the files based on the similarity between the file-level reasoning and the bug report:
\begin{tcolorbox}[colback=gray!5!white, colframe=black!75!white, boxrule=0.5pt, arc=2pt, fontupper=\ttfamily\footnotesize, left=2pt, right=2pt, top=2pt, bottom=2pt]
Below is a list of files from a repository and the reasoning behind the code of these files:
\{file\_reasoning\}

Can you rank the files based on the similarity of their reasoning to the bug report:
\{bug\_report\}

Please just return the list of ranked files.
\end{tcolorbox}

At this stage, we also utilize an embedding model to rank the files based on their reasoning and compare the results with those obtained from LLM-based ranking. We choose the better approach for the next step.
    
\noindent\textbf{Element-level Localization:} Not every part of a file needs to be changed. So, for each of the top-ranked files, we perform element-level localization by generating reasoning for each function, class, or global variable (5), and ranking them based on how relevant they seem to the bug report (6).\\

\noindent\textit{Element-Level Reasoning Prompt:}
After identifying top-k relevant files, we extract all functions, classes, and global variables and ask the LLM to explain their functionality with respect to the bug report:
\begin{tcolorbox}[colback=gray!5!white, colframe=black!75!white, boxrule=0.5pt, arc=2pt, fontupper=\ttfamily\footnotesize, left=2pt, right=2pt, top=2pt, bottom=2pt]
A user is trying to fix a bug described in the following report:
\{bug\_report\}

Below is a code element (a function, a class, or a global variable) in a file in a repository:
\{element\}

Explain the purpose and functionality of this code element in the context of the bug report.
Focus on what this element does and whether it may be related to the bug.
\end{tcolorbox}
\noindent\textit{Element Ranking Prompt:}
We then ask the LLM to identify the most relevant code elements by comparing their reasoning to the bug report:
\begin{tcolorbox}[colback=gray!5!white, colframe=black!75!white, boxrule=0.5pt, arc=2pt, fontupper=\ttfamily\footnotesize, left=2pt, right=2pt, top=2pt, bottom=2pt]
You are provided with a list of code elements (functions, classes, and global variables) from a repository, along with an explanation of what each element does:
\{file\_elements\_reasoning\}

Also, you are given the following bug report:
\{bug\_report\}

Based on the reasoning for each code element and the bug report, which of these elements are most likely related to the bug?
Please just return a ranked list of the potentially buggy elements (keys in the file\_elements\_reasoning dictionary) without any further explanation.
\end{tcolorbox}

\noindent\textbf{Line Localization and repair:}
We intentionally do not introduce reasoning at the line level. First, line-level reasoning is inherently ill-posed when lines are considered in isolation: individual lines often lack sufficient semantic context, and explaining them independently (without surrounding control flow, data flow, or neighboring statements) is unlikely to yield meaningful or reliable explanations.

Second, our reasoning framework is designed to operate on a single candidate at a time (file or code element) rather than over the entire repository. Extending this paradigm to the line level would require prompting the LLM separately for every line in the candidate files. Given that real-world repositories contain thousands of lines per instance, this approach would be computationally prohibitive, economically impractical, and unlikely to improve localization accuracy proportionally. 

For these reasons, we keep the line localization and repair stages unchanged and follow the standard Agentless pipeline. This choice isolates the effect of our reasoning-guided modifications at the file and element levels, ensuring that any observed gains can be attributed to improved higher-level localization rather than to changes in low-level heuristics. So, we provide the full code of the top-ranked elements to the LLM and ask it to narrow down the list to the most likely places that need to be edited — whether that's a class, a function, or even specific lines of code(7). For the repair phase, the LLM is provided with the bug report together with the localized code fragments (i.e., the top-ranked edit locations identified in earlier steps). The model is prompted to propose multiple candidate patches, each representing a plausible fix for the reported issue. Generating multiple candidates is important because the first attempt is often incomplete or imprecise, and diversity increases the chance of producing a correct repair(8). Once the candidate patches are generated, they are passed to the patch validation module, which executes the modified program against the project’s regression test suite to help rank and choose the best patch(9). Finally, the system selects the highest-ranked passing patch as the final output.

\section{Experiment Settings}

\subsection{Dataset and Model}

\subsubsection{Dataset}
We conduct our experiments on multiple subsets of the SWE-bench benchmark, which is designed to evaluate large language models on real-world software engineering tasks by generating patches that resolve GitHub issues within a given codebase. Our experiments include:\\

\noindent\textbf{SWE-bench Verified:\cite{jimenez2023swe, swebench_verified_dataset}} 
A curated subset developed in collaboration with OpenAI Preparedness. SWE-bench Verified consists of 500 issue instances that have been manually reviewed and confirmed by experienced software engineers to be solvable. Each instance includes:
A natural language issue description (\texttt{problem\_statement}),
A snapshot of the target repository (\texttt{repo}, \texttt{base\_commit}),
A human-authored patch and associated test changes (\texttt{patch}, \texttt{test\_patch}).
This subset is well-suited for evaluating both FL and APR with high confidence.

\noindent\textbf{SWE-bench Lite:\cite{jimenez2023swe, swebench_lite_dataset}} 
A smaller subset of SWE-bench containing 300 issue instances. Lite is designed to provide a quicker evaluation of APR methods while still maintaining a representative variety of software bugs across different repositories. Each instance includes the same fields as Verified, allowing consistent comparison of FL and repair performance, but at a lower computational cost.

\noindent\textbf{SWE-bench Java:\cite{zan2024swe, zan2025multiswebench}} 
A language-specific subset focusing on Java projects, containing 91 issue instances. SWE-bench Java enables the evaluation of APR and FL approaches on a language other than Python, providing insights into cross-language generalizability. Each instance includes a Java repository snapshot, a natural language bug description, and human-authored patches and tests.

\subsubsection{LLM}
For our FL pipeline, we employed different reasoning models, including Gemini 2.5 Pro, Claude 4 Sonnet, and o4-mini. These models were selected for their ability to understand and reason over source code in the context of natural language bug reports.\\

\noindent\textbf{Gemini 2.5 Pro:\cite{comanici2025gemini}} A high-capacity model optimized for reasoning about code and complex problem statements. It provides strong performance in understanding code structure and generating relevant explanations.

\noindent\textbf{Claude 4 Sonnet:\cite{claude4sonnet}} Designed for general-purpose reasoning, Claude 4 Sonnet demonstrates high reliability in producing structured, interpretable outputs, which is beneficial for both file-level and element-level localization.

\noindent\textbf{o4-mini:\cite{o3o4mini}} A lightweight, efficient model suitable for fast reasoning tasks. While smaller than the other models, o4-mini is cost-effective and can still generate meaningful reasoning for code elements.

We first evaluated all three LLMs on our file localization tasks to determine which one produced the most accurate reasoning for ranking files. Due to cost constraints, the best-performing LLM was then selected for three purposes:

\begin{enumerate}
    \item Generating reasoning and ranking in the element localization step.
    \item Comparison with baselines in fault localization – To assess the improvements provided by reasoning-based ranking, we compared our top LLM against baseline methods.
    \item Comparison with baselines in APR – The same top-performing model was used to guide the repair phase, allowing us to evaluate how reasoning-guided localization can influence the overall patch generation quality.
\end{enumerate}

\textbf{Model Choice:} For internal comparisons between Agentless and RGFL, we always use the same LLM (Gemini 2.5 Pro). For external agents (OpenHands, OrcaLoca, AutoCodeRover), we report their best publicly reported configuration from the SWE-bench leaderboard, which uses different LLMs (e.g., Claude variants).\\

Furthermore, for the embedding-based approach in the file localization step, to rank the files based on the embedding, we use gemini-embedding-001\cite{google_gemini_embedding}, Voyage 3.5\cite{voyage-3.5-blog-2025}, text-embedding-3-small\cite{openai_text_embedding_3_small} to embed both bug reports and reasoning outputs, enabling cosine-similarity-based ranking of candidate files.\\

\noindent\textbf{Gemini-embedding-001:} It is designed to support both natural language and code retrieval tasks. It is particularly effective for scenarios that combine bug reports written in natural language with reasoning generated over source code, since it captures cross-modal semantic similarity. 

\noindent\textbf{Voyage 3.5:} Provides specialized embedding models optimized for code retrieval and long-context technical text. Voyage 3.5 consistently achieves state-of-the-art performance on public code search and retrieval benchmarks, making it a strong choice for fault localization. Since Anthropic (Claude) does not provide standalone embedding models, Voyage fills this gap by offering robust embeddings that can be paired with any LLM. We therefore employ Voyage 3.5 as a complementary alternative to Gemini and OpenAI embeddings.

\noindent\textbf{Text-embedding-3-small:}  OpenAI’s text-embedding-3-small is a cost-efficient, widely adopted embedding model. Although it is not as specialized for code as Voyage, it serves as a reproducible baseline since it has been widely used in prior work on IR-based fault localization. Including this model in our evaluation provides a fair reference point for measuring improvements from more advanced or specialized embeddings.

\subsection{Evaluation metrics}
To evaluate the effectiveness of our \textit{file localization} approach, we use the following standard retrieval metrics:\\

\noindent\textbf{Hit@k\cite{manning2008introduction}:} Measures whether the ground truth file is present within the top-k retrieved files. Formally, Hit@k = 1 if the correct file is among the top k candidates; otherwise, it is 0. We report the average Hit@k over all instances. This metric captures how often the localization system successfully retrieves the relevant file within a limited number of guesses.

\noindent\textbf{Recall@k\cite{manning2008introduction}:} Measures the fraction of all relevant files that are retrieved in the top-k results. For single-label FL, Recall@k is numerically the same as Hit@k, but in multi-label settings, it generalizes to more than one relevant file. It reflects the completeness of the retrieved set.

\noindent\textbf{MRR (Mean Reciprocal Rank\cite{craswell2009mrr}):} Computes the average of the reciprocal ranks of the first relevant file across all instances. If the ground truth file is ranked at position r, its reciprocal rank is 1/r. This metric rewards systems that not only retrieve the correct file but also rank it higher in the list. A higher MRR indicates that relevant files tend to appear near the top of the ranking.

To evaluate the effectiveness of our \textit{element localization} approach within the top-k retrieved files, we use the following metrics:\\

\noindent\textbf{Exact Match:} Measures the proportion of instances where the predicted set of relevant elements (e.g., classes or functions) exactly matches the annotated ground truth. This metric is strict and only rewards completely correct predictions.\\

\noindent Finally, to evaluate the end-to-end \textit{repair} effectiveness, we report:\\

\noindent\textbf{Resolved Rate\cite{jimenez2023swe}:} The percentage of instances in which the system successfully applies a patch that passes tests. This reflects the ultimate goal of automated program repair—whether the system can produce a valid fix that resolves the issue.

\subsection{Experimental setup}

We integrate our LLM-based reasoning approach into the modular APR framework—Agentless, changing both the file-level and element-level localization modules. Our pipeline proceeds in four stages: (1) file localization using code+reasoning ranking, (2) element localization using reasoning within the top-k files (which we use k=3 in our study),(3) line localization, and (4) patch generation and validation.

In the repair phase, consistent with the Agentless pipeline, we generate 10 samples per instance, repeated across 4 independent runs, resulting in 40 candidate patches per instance. These patches are then passed through the patch validation and selection module, which evaluates them against the existing regression tests to identify successful repairs.

The average cost of RGFL per sample is around \$4.4, so it incurs a higher computational cost than Agentless (3.7×). However, the resulting improvement in fault localization and repair accuracy is substantial. Given that end-to-end repair performance is highly sensitive to localization quality, the +12.8\% absolute increase in resolved rate provides a compelling justification for the added cost when accuracy is the priority.

\subsection{Research Questions}
To evaluate the effectiveness of LLM-based reasoning in APR, we address the following research questions:\\
\noindent\textbf{RQ1: Can LLM-generated reasoning about source code files and elements improve file localization and element localization performance in a modular APR system like Agentless?}

\textbf{RQ1.1:} First, we examine whether re-ranking files based on their reasoning relevance improves the inclusion and rank of found files. Since we first retrieve candidate files using code-based similarity and then re-rank them using LLM-generated reasoning, our approach effectively leverages both the source code and its associated reasoning to identify the most relevant files. 
    
For this RQ, we explore two complementary strategies: (1) embedding-based re-ranking, where both the bug report and generated reasoning are embedded and compared using cosine similarity, and (2) direct LLM-based re-ranking, where the LLM itself is prompted to assess reasoning relevance to the bug report. The latter approach addresses potential weaknesses in embedding-only methods, as LLM judgments may capture deeper semantic connections.

For this RQ, we used different LLMs like Gemini 2.5 pro, Claude 4 Sonnet, and o4-mini to find which one works better for our approach.
    
\textbf{RQ1.2:} We test whether reasoning-based element ranking increases the likelihood of identifying correct elements(functions, classes, and global variables) compared to code-based ranking. For that, we compared three setups: (1) Agentless file + Agentless element localization, (2) RGFL file localization + Agentless element localization, and (3) RGFL file localization + RGFL element localization. For this experiment, we employ our strongest-performing LLM from the previous step (Gemini 2.5 Pro). In addition, to contextualize the achievable performance, we compute an upper bound for element-level localization based on the results of file-level localization, providing insight into how much improvement can be gained beyond file identification.\\

\noindent\textbf{RQ2: How generalizable is our fault localization approach across different datasets and programming languages, and how does it compare to existing baseline methods?} 

In this research question, we compare RGFL against several baselines across multiple datasets. For baselines, we include a simple RAG method with BM25 retrieval, as well as state-of-the-art agents such as Agentless, OpenHands, AutoCodeRover, and OrcaLoca. We evaluate on SWE-bench Verified, SWE-bench Lite, and SWE-bench Java to test the generalizability of our method across different datasets and programming languages.

\textbf{RQ2.1:} We want to see if our File localization approach is better than other baselines in different datasets and languages, like Java, in addition to Python. 

\textbf{RQ2.2:} We want to see if our Element localization approach is better than other baselines in different datasets and languages, like Java, in addition to Python. 
For this part, we used the best reported LLMs of other baselines and compared them with the results of our approach with our best LLM (Gemini 2.5 pro).\\
 
\noindent\textbf{RQ3: Does ranking files and elements based on LLM reasoning improve the success rate of final program repairs compared to baseline agents?}

we evaluate whether enhancing the file localization and element localization stages with LLM-generated reasoning improves the overall success rate of APR. To do this, we modified the Agentless pipeline in two ways while leaving line localization and patch generation unchanged: (1) Replacing only the file localization component with our reasoning-based approach. (2) Replacing both file localization and element localization with our reasoning-based methods. This setup allows us to isolate the impact of each stage on final repair outcomes while ensuring a fair comparison with the original Agentless system.
    
Furthermore, for unresolved cases, we categorize errors into (a) file-level miss (ground-truth files not in top-k), (b) element-level miss (ground-truth element not found), (c) line-level miss (couldn't find the exact buggy line), and (d) repair-level miss (the buggy location found, but the repair is unsuccessful). We then compute counterfactual upper bounds by injecting ground-truth at each level (file, element, line) to quantify how much of the remaining gap is attributable to fault localization versus patch generation.

\section{Experiment Results}

\subsection{RQ1: Can LLM-generated reasoning about source code files and elements improve file localization and element localization performance in a modular APR system like Agentless?}

%\textbf{RQ1.1:} Incorporating LLM-generated reasoning at the file level improved the ordering of candidate files produced by Agentless. By re-ranking these candidates based on their inferred relevance to the bug report, the ground truth file was more likely to appear near the top of the list. While the total set of retrieved files remained the same, leading to no change in the Hit@5 value (most of the instances contain fewer than five retrieved files in total), the refined ranking still produced consistent accuracy gains across evaluation metrics.

\textbf{RQ1.1:} As shown in Table~\ref{tab:file_localization}, RGFL consistently improves file localization across all LLMs. The most notable gains appear at small k values: for Hit@1, Gemini 2.5 Pro improves from 71.4\% to 85\% (19.05\% improvement), Claude 4 Sonnet from 75.4\% to 83.6\% (10.88\% improvement), and o4-mini from 68.4\% to 80.8\% (18.13\% improvement). Recall@1 shows similar improvements of 10-19\% across models, while MRR improves by 6–10\% (e.g., Gemini from 81.8\% to 88.8\%, Claude from 83.9\% to 88.5\%, o4-mini from 77.2\% to 84.7\%), indicating that ground-truth files are ranked closer to the top on average. Improvements flatten at larger k values, with Hit@5 converging around 93–94\%, since Agentless already retrieves the correct file(s) within that range. These gains are especially impactful because the Agentless pipeline depends on the top-3 ranked files to guide element-level localization and subsequent patch generation. Ensuring that the ground-truth file enters this top-3 window directly increases the likelihood of successful downstream repair.

\begin{figure*}[t]
    \centering
    \begin{subfigure}{0.32\linewidth}
        \centering
        \includegraphics[width=\linewidth]{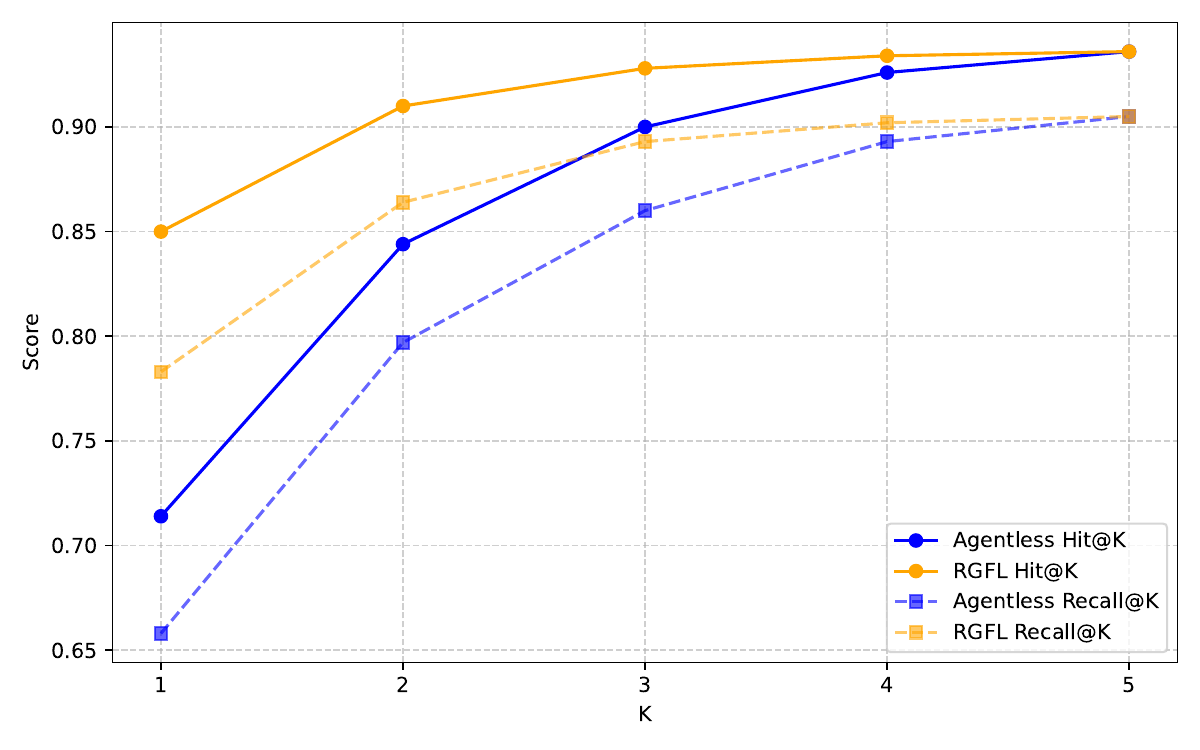}
        \caption{Gemini 2.5 Pro}
        \label{fig:gemini_rq1}
    \end{subfigure}
    \hfill
    \begin{subfigure}{0.32\linewidth}
        \centering
        \includegraphics[width=\linewidth]{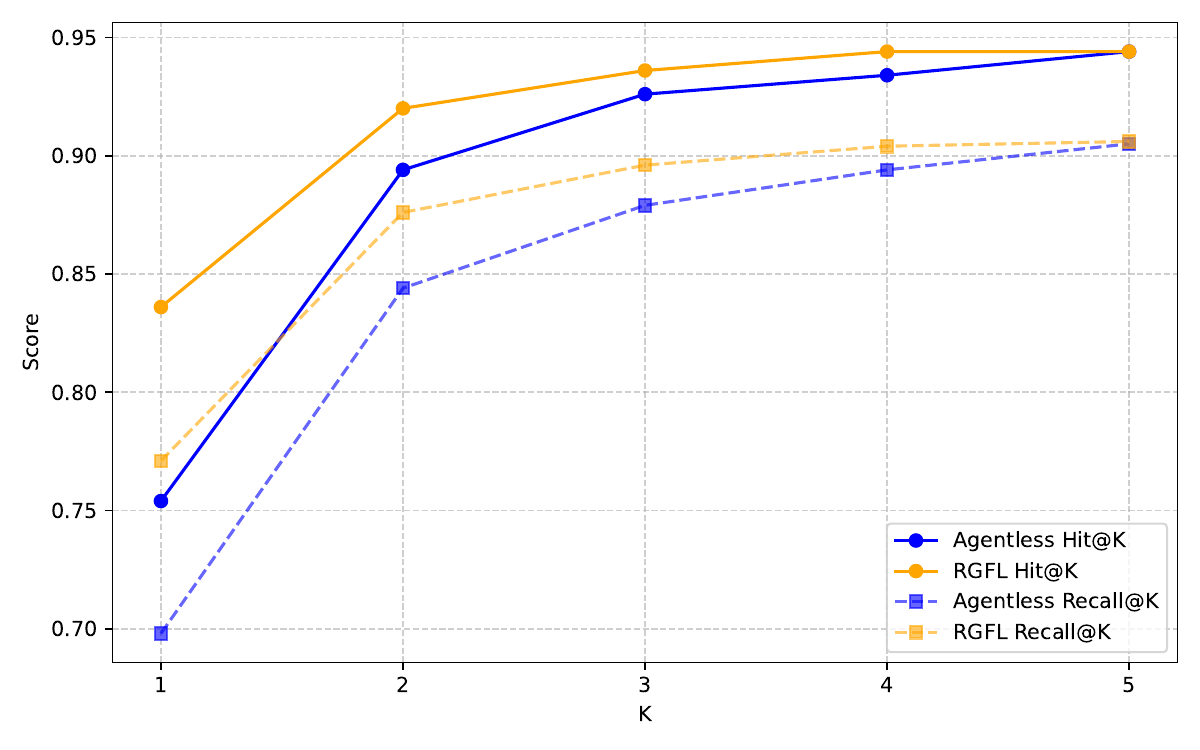}
        \caption{Claude 4 Sonnet}
        \label{fig:claude4_rq1}
    \end{subfigure}
    \hfill
    \begin{subfigure}{0.32\linewidth}
        \centering
        \includegraphics[width=\linewidth]{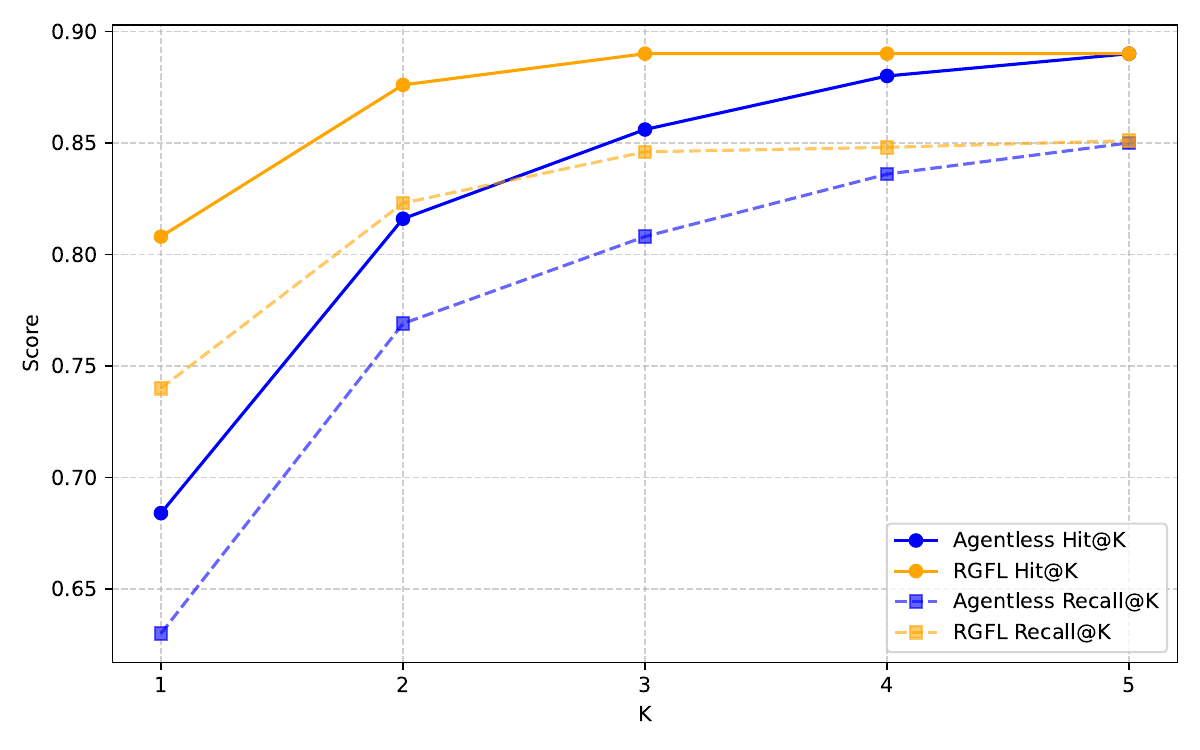}
        \caption{GPT-4o mini}
        \label{fig:o4_rq1}
    \end{subfigure}
    \caption{File localization performance (Hit@k, Recall@k) with and without LLM-based reasoning on the SWE-bench Verified dataset.}
    \label{fig:fig2}
\end{figure*}

Figure~\ref{fig:fig2} further illustrates these improvements through Hit@k and Recall@k curves, showing how reasoning-based reranking provides consistent benefits across different cutoff values.

A natural question is whether the same improvements could be achieved more cheaply using embedding models on the generated reasoning instead of LLM-based reranking. Table~\ref{tab:file_localization} shows that this approach does not succeed. Embedding-based reranking either underperforms or fails to improve over the Agentless baseline (e.g., Hit@1 falls to 64.4\% with Gemini-embedding-001, to 60.8\% with Voyage 3.5, and to 52\% with text-embedding-3-small). The limitation is that embeddings compress bug reports and explanations into fixed vectors, capturing broad semantic similarity but missing the fine-grained, task-specific signals required for accurate FL. In contrast, when the same LLM both generates and ranks reasoning, it can apply deeper contextual understanding to identify the truly relevant files. These results demonstrate that natural language reasoning itself—not just its embedding representation—is essential for consistent improvements.

\begin{table*}[t!]
\centering
\caption{File localization performance (Hit@k, Recall@k, and MRR) with and without LLM-based reasoning on the SWE-bench Verified dataset.}
\label{tab:file_localization}
\begin{adjustbox}{max width=\textwidth}
\begin{tabular}{@{}lccccc|ccccc|c@{}}
\rowcolor{gray!20}
\textbf{Method} & \multicolumn{5}{c|}{\textbf{Hit@k}} & \multicolumn{5}{c|}{\textbf{Recall@k}} & \textbf{MRR} \\
\cmidrule(lr){2-6}\cmidrule(lr){7-11}
 & k=1 & k=2 & k=3 & k=4 & k=5 & k=1 & k=2 & k=3 & k=4 & k=5 &  \\
\toprule
Agentless (Gemini 2.5 Pro) & 71.4\% & 84.4\% & 90\% & 92.6\% & 93.6\% & 65.8\% & 79.7\% & 86\% & 89.3\% & 90.5\% & 81.8\% \\
RGFL + Embedding (Gemini-embedding-001) & 64.4\% & 84.2\% & 89.8\% & 92.4\% & 93.2\% & 59\% & 79\% & 85.8\% & 88.7\% & 89.9\% & 77\% \\
RGFL (Gemini 2.5 Pro) & \textbf{85\%} & \textbf{91\%} & \textbf{92.8\%} & \textbf{93.4\%} & \textbf{93.6\%} & \textbf{78.3\%} & \textbf{86.4\%} & \textbf{89.3\%} & \textbf{90.2\%} & \textbf{90.5\%} & \textbf{88.8\%} \\
\midrule
Agentless (Claude 4 Sonnet) & 75.4\% & 89.4\% & 92.6\% & 93.4\% & 94.4\% & 69.8\% & 84.4\% & 87.9\% & 89.4\% & 90.5\% & 83.9\% \\
RGFL + Embedding (Voyage 3.5) & 60.8\% & 83.2\% & 91.8\% & 94.2\% & 94.4\% & 55.3\% & 77.7\% & 87.7\% & 90.4\% & 90.6\% & 75.5\% \\
RGFL (Claude 4 Sonnet) & \textbf{83.6\%} & \textbf{92\%} & \textbf{93.6\%} & \textbf{94.4\%} & \textbf{94.4\%} & \textbf{77.1\%} & \textbf{87.6\%} & \textbf{89.6\%} & \textbf{90.4\%} & \textbf{90.6\%} & \textbf{88.5\%} \\
\midrule
Agentless (o4-mini) & 68.4\% & 81.6\% & 85.6\% & 88\% & 89\% & 63\% & 76.9\% & 80.8\% & 83.6\% & 85\% & 77.2\% \\
RGFL + Embedding (text-embedding-3-small) & 52\% & 72.6\% & 84\% & 87.4\% & 88.6\% & 46.8\% & 67.2\% & 79.3\% & 83.2\% & 84.5\% & 67.3\% \\
RGFL (o4-mini) & \textbf{80.8\%} & \textbf{87.6\%} & \textbf{89\%} & \textbf{89\%} & \textbf{89\%} & \textbf{74\%} & \textbf{82.3\%} & \textbf{84.6\%} & \textbf{84.8\%} & \textbf{85.1\%} & \textbf{84.7\%} \\
\bottomrule
\end{tabular}
\end{adjustbox}
\end{table*}

Based on these results, we adopt LLM-based ranking for our approach rather than embedding-based ranking, since it consistently yields stronger improvements.

For the next stage (element-level localization), we needed to select the most effective LLM to serve as the backbone of our approach. Both Claude 4 Sonnet and Gemini 2.5 Pro achieve nearly identical performance in file localization: the average difference across Hit@k and Recall@k between the two models is only $\approx 0.1\%$, which is negligible. Given this parity in accuracy, cost becomes the deciding factor. Running element-level reasoning with Claude would cost around \$2,000 for the SWE-bench Verified dataset, whereas Gemini could do the job on Google Cloud Vertex AI with free credits, approximately a worth of \$400 . This makes Gemini a substantially more cost-effective option, allowing us to scale experiments without sacrificing performance.

For element-level localization, we follow Agentless and pass only the top-3 files from the file-ranking stage. This is because the vast majority of bugs modify a single file (429 out of 500 instances), and almost all remaining cases involve less than 4 files. Only three outliers touch 5, 6, and 21 files. Consequently, a top-3 cutoff captures the ground-truth file in nearly all instances while keeping the element search space small—reducing distractor context and computation without sacrificing recall.
% Figure \ref{fig:fig3} supports this choice: the vast majority of bugs modify a single file (429 instances), and almost all remaining cases involve less than 4 files. Only three outliers touch 5, 6, and 21 files. Consequently, a top-3 cutoff captures the ground-truth file in nearly all instances while keeping the element search space small—reducing distractor context and compute without sacrificing recall.
\begin{comment}
\begin{figure}
    \centering
    \includegraphics[width=0.7\linewidth]{figures/files_per_instance.pdf}
    \caption{Histogram of Files Changed per Instance}
    \label{fig:fig3}
\end{figure}
\end{comment}

%This research question investigates whether reasoning-based ranking of code elements (i.e., functions, classes, and global variables) can enhance the element-level fault localization stage. Rather than relying on the Agentless approach, which ranks elements based on code similarity to the bug report, we generate natural language reasoning for each code element, guided by a bug-specific prompt, and ask the LLM to rank elements based on their semantic relevance to the bug.

%To evaluate this, we first apply our file-level localization method, which uses reasoning-based ranking. Then, we compare three setups:
\begin{comment}
    \begin{itemize}
    \item Agentless file + Agentless element localization
    \item RGFL file localization + Agentless element localization
    \item RGFL file localization + RGFL reasoning-based element localization
    \end{itemize}
\end{comment}
\begin{table}[t]
\centering
\caption{Element localization (Exact Match) with and without LLM-based reasoning on SWE-bench Verified.}
\label{tab:element_localization}
\begin{adjustbox}{max width=0.9\columnwidth}
\begin{tabular}{@{}l l c c@{}}
\toprule
\rowcolor{gray!20}
\textbf{File Localization Method} & \textbf{Element Localization Method} & \textbf{Exact Match} & \textbf{Upper bound} \\
\midrule
Agentless (Top-3 files) & Agentless & 36\% & 70\% \\
RGFL (Top-3 files)      & Agentless & 41\% & 80\% \\
RGFL (Top-3 files)      & RGFL      & \textbf{69\%} & 80\% \\
\bottomrule
\end{tabular}
\end{adjustbox}
\end{table}
\textbf{RQ1.2:} Table~\ref{tab:element_localization} reports element-level exact match on SWE-bench Verified. Using Agentless’s own file localization, element localization reaches 36\%, while the upper bound is 70\%. When we replace the file localization step with RGFL, the same Agentless element localizer improves to 41\% (13.89\% improvement), with a higher upper bound of 80\%. Finally, applying reasoning at both the file and element level yields the largest gain: exact match increases to 69\%, which is 91.67\% improvement over Agentless and 68.29\% improvement over using RGFL only at the file stage. 

%The upper bound for element-level localization is defined by how often all ground-truth buggy files appear within the top-K ranked files. Only instances where every buggy file is retrieved are “covered,” and the fraction of such cases gives the maximum achievable exact match. This reflects the best possible element localization assuming a perfect ranker, with performance limited only by file retrieval accuracy.
The upper bound for element-level localization is determined by file coverage. For each instance, if all ground-truth files appear within the top-k retrieved files, we count the instance as "covered". The fraction of covered instances across the dataset gives the maximum achievable element-level exact match, assuming a perfect element ranker. This reflects the fact that element localization can only succeed if the necessary buggy files are first retrieved.

In summary, we can say: (a) reasoning-based file localization directly improves downstream element localization, and (b) reasoning-based element localization provides an additional boost, nearly closing the gap to the theoretical upper bound. %This demonstrates that LLM reasoning is crucial for element-level understanding in modular APR systems.

\mybox{\textbf{Answer to RQ1:} LLM-generated reasoning improves both file and element localization in modular APR systems like Agentless. At the file level, reranking with reasoning consistently ranked the ground-truth file higher—especially within the top-3 files used for the downstream repair—whereas embedding-based reranking often fails due to shallow similarity matches. At the element level, combining reasoning-based file and element localization leads to a large jump in exact match accuracy (from 36\% with Agentless to 69\%), showing that LLM reasoning provides deeper semantic alignment between bug reports and code.}

\subsection{RQ2: How generalizable is our fault localization approach across different datasets and programming languages, and how does it compare to existing baseline methods?}

\textbf{RQ2.1:} In this research question, we compare our file localization approach against several baselines across multiple datasets. Table~\ref{tab:cross_dataset_file_localization} reports the results. For each baseline, we used their best-reported model from the dataset leaderboard. The only exceptions are: (1) for Agentless, since it serves as the foundation of our approach, we used the same model as ours (Gemini 2.5 Pro), and (2) for OrcaLoca, whose reported best model was Claude 3.5 Sonnet, we instead ran Claude 4 Sonnet (a better Claude model), since their trajectory data was not publicly available.

\begin{figure*}[t]
    \centering
    \begin{subfigure}{0.32\linewidth}
        \centering
        \includegraphics[width=\linewidth]{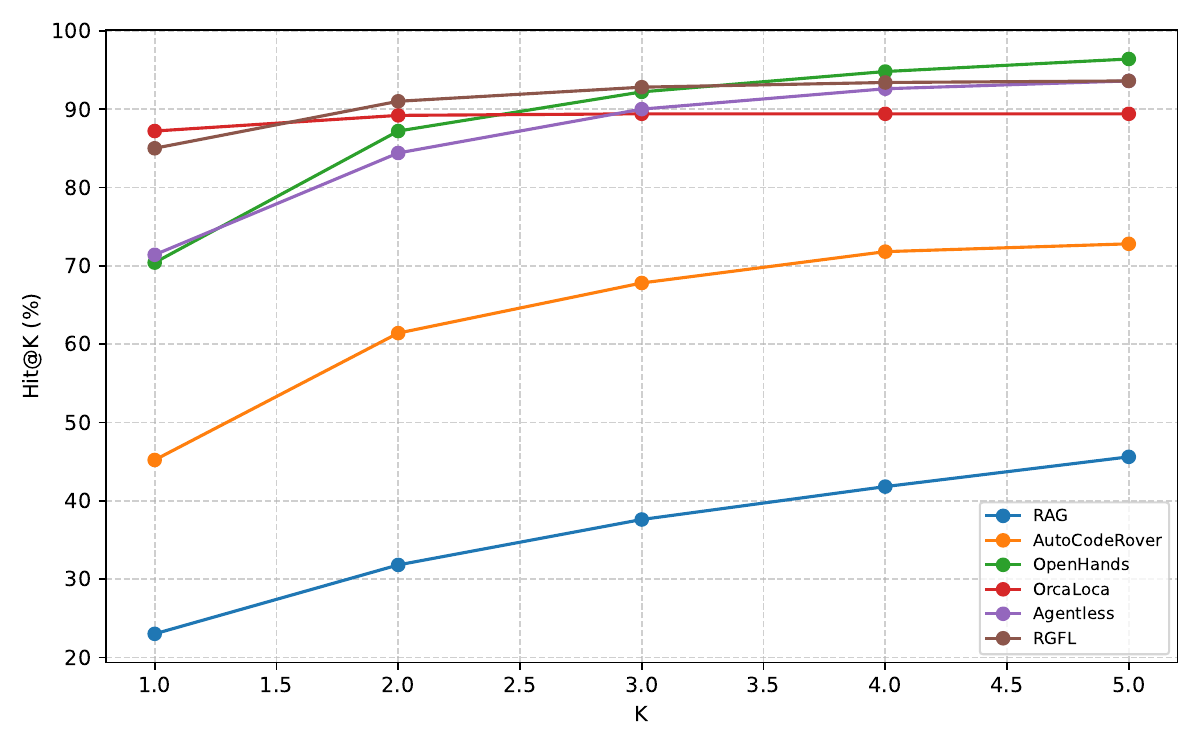}
        \caption{SWE-bench Verified}
        \label{fig:verified_rq2}
    \end{subfigure}
    \hfill
    \begin{subfigure}{0.32\linewidth}
        \centering
        \includegraphics[width=\linewidth]{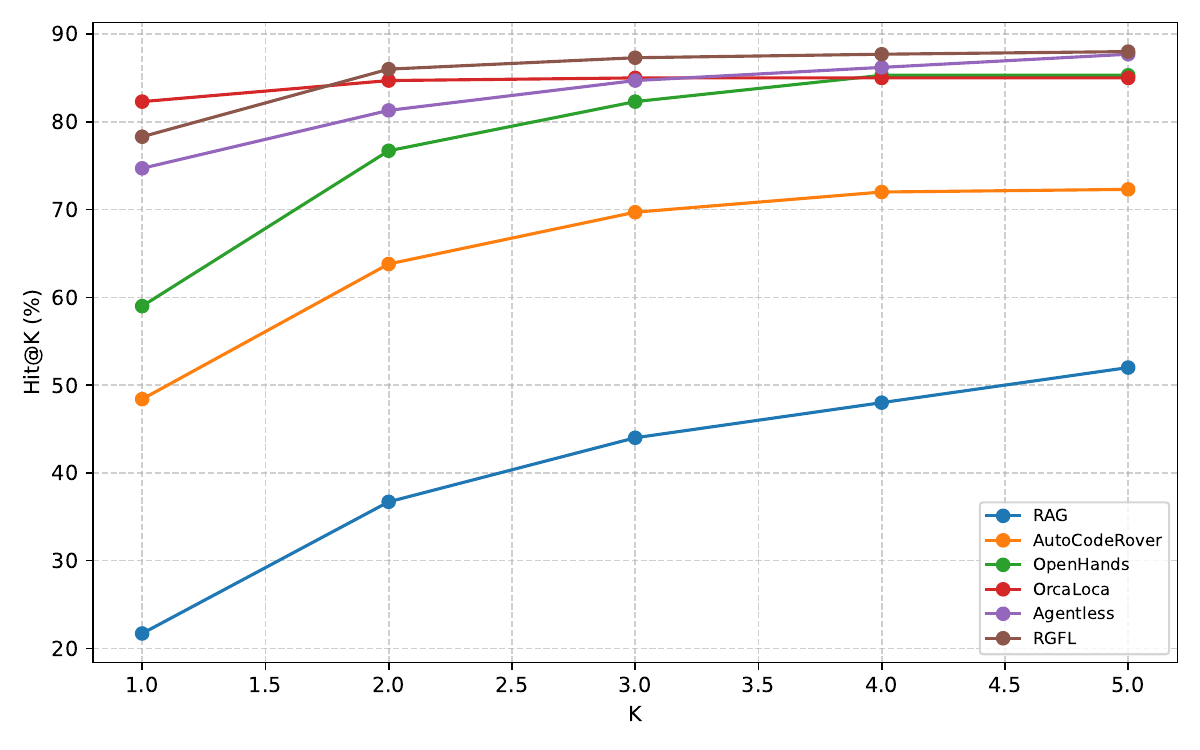}
        \caption{SWE-bench Lite}
        \label{fig:lite_rq2}
    \end{subfigure}
    \hfill
    \begin{subfigure}{0.32\linewidth}
        \centering
        \includegraphics[width=\linewidth]{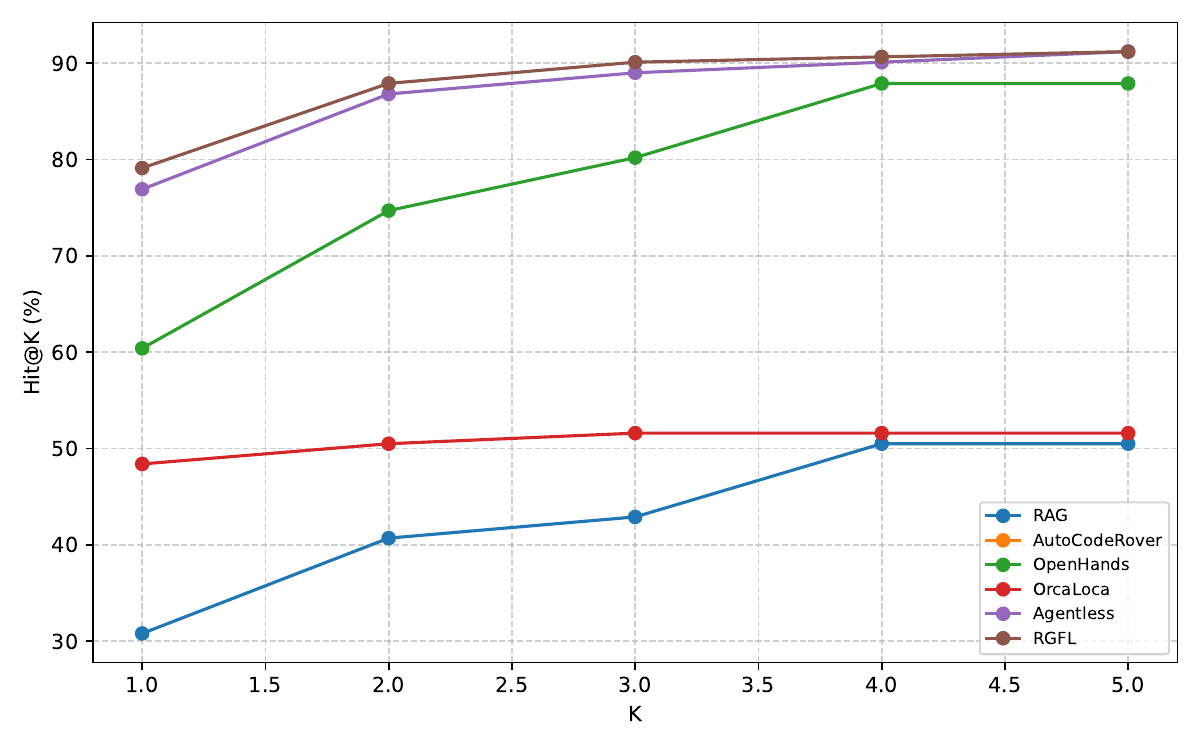}
        \caption{SWE-bench Java}
        \label{fig:java_rq2}
    \end{subfigure}
    \caption{Cross-dataset comparison of file localization performance (Hit@k) across different methods.}
    \label{fig:fig3}
\end{figure*}

The results in Table~\ref{tab:cross_dataset_file_localization} show that our method generally outperforms other approaches. Figure~\ref{fig:fig3} further illustrates this through Hit@k plots across datasets. 

Looking at SWE-bench Verified, OrcaLoca performs slightly better at Hit@1 and Recall@1, indicating that its first retrieved file is often correct. However, its advantage diminishes at larger k values, where our method achieves stronger performance. OpenHands also performs competitively beyond Hit@3, but since our approach re-ranks files produced by Agentless, our maximum achievable performance is bounded by Agentless’s top-5 retrieval (93.6\%). Because we pass only the top-3 files to the next stage (element-level localization), for all methods to have a fair comparison, we only reported Hit@3 and Recall@3 in the table. Our method achieves the best Hit@3, Recall@3, and MRR, indicating superior overall ranking quality.

On SWE-bench Lite, the same pattern emerges. OrcaLoca again leads at Hit@1 and Recall@1, while our method consistently outperforms at larger k values. In this dataset, unlike in Verified, OpenHands (Claude 3.5 Sonnet) does not surpass our approach, which may be due to model differences (Claude 3.5 vs. Claude 4 Sonnet in Verified). Additionally, note that in this dataset, Hit@k and Recall@k values are identical because each instance in this dataset has only a single ground-truth file.

On SWE-bench Java, our method clearly achieves the best performance across all metrics. We excluded AutoCodeRover because its agent is not designed for Java; despite attempts, we could not adapt their code to this dataset. OrcaLoca also performed much worse compared to its results on other datasets, likely because its implementation was not designed for Java either. By contrast, our method maintained strong results consistent with its performance on Python datasets, demonstrating its robustness and generalizability across different datasets and programming languages.\\
\begin{table*}[t!]
\centering
\caption{Cross-dataset comparison of file localization performance (Hit@3, Recall@3, and MRR) across different agents and models on SWE-bench Verified, SWE-bench Lite, and SWE-bench Java datasets.}
\label{tab:cross_dataset_file_localization}
\begin{adjustbox}{max width=0.75\textwidth}
\begin{tabular}{@{}lllccc@{}}
\rowcolor{gray!20}
\textbf{Dataset} & \textbf{Agent} & \textbf{Model} & \textbf{Hit@3} & \textbf{Recall@3} & \textbf{MRR} \\
\toprule
\multirow{6}{*}{Verified} 
& RAG & -- & 37.6\% & 34.7\% & 32.5\% \\
& AutoCodeRover & (Claude 3.5 Sonnet) & 67.8\% & 63.2\% & 57\% \\
& OpenHands & (Claude 4 Sonnet) & 92.2\% & 88\% & 81.7\% \\
& OrcaLoca & (Claude 4 Sonnet) & 89.4\% & 83.1\% & 88.3\% \\
& Agentless & (Gemini 2.5 Pro) & 90\% & 86\% & 81.8\% \\
& RGFL & (Gemini 2.5 Pro) & \textbf{92.8\%} & \textbf{89.3\%} & \textbf{88.8\%} \\
\midrule
\multirow{6}{*}{Lite} 
& RAG & -- & 44\% & 44\% & 31\% \\
& AutoCodeRover & (GPT-4o) & 69.7\% & 69.7\% & 60.2\% \\
& OpenHands & (Claude 3.5 Sonnet) & 82.3\% & 82.3\% & 70.5\% \\
& OrcaLoca & (Claude 4 Sonnet) & 85\% & 85\% & \textbf{83.6\%} \\
& Agentless & (Gemini 2.5 Pro) & 84.7\% & 84.7\% & 79.9\% \\
& RGFL & (Gemini 2.5 Pro) & \textbf{87.3\%} & \textbf{87.3\%} & 82.8\% \\
\midrule
\multirow{5}{*}{Java} 
& RAG & -- & 42.9\% & 27.6\% & 39.9\% \\
& OpenHands & (Claude 3.7 Sonnet) & 80.2\% & 51.3\% & 71.8\% \\
& OrcaLoca & (Claude 4 Sonnet) & 51.6\% & 28.4\% & 49.8\% \\
& Agentless & (Gemini 2.5 Pro) & 89\% & 56.8\% & 83.3\% \\
& RGFL & (Gemini 2.5 Pro) & \textbf{90.1\%} & \textbf{57.6\%} & \textbf{84.5\%} \\
\bottomrule
\end{tabular}
\end{adjustbox}
\end{table*}
\textbf{RQ2.2:} In this research question, we compare the element-level localization performance of our method against the baselines introduced in RQ2.1, across multiple datasets. Table~\ref{tab:element_localization_baseline} reports the Exact Match scores for each method, where all approaches use the top-3 files obtained from the file localization stage.

Our method achieves the best performance on SWE-bench Verified(69\%) and SWE-bench Java(46\%), with OpenHands as the next strongest baseline (64\% and 38\% respectively). The performance gap of 5 points on Verified and 8 points on Java shows that our improvements are consistent across datasets, even when baselines are competitive. On SWE-bench Lite, our method is only slightly lower (68\%) than OpenHands (71\%), a gap of about 3 points, which is reasonable given that OpenHands with Claude currently ranks as the strongest method on the official leaderboard. OrcaLoca delivers moderate performance on Verified and Lite, but performs poorly on Java, which is expected since its design is not tailored to non-Python datasets. Overall, these results demonstrate that our method generalizes well to different datasets and languages, while maintaining strong performance relative to competitive baselines.

\begin{table*}[t!]
\centering
\caption{Element-level localization performance (Exact Match) across datasets and agents, using top-3 files as input.}
\label{tab:element_localization_baseline}
\begin{adjustbox}{max width=0.75\textwidth}
\begin{tabular}{@{}lll r@{}}
\rowcolor{gray!20}
\textbf{Dataset} & \textbf{Agent} & \textbf{Model} & \textbf{Exact Match (\%)} \\
\toprule
\multirow{6}{*}{Verified} 
& RAG & (BM25) & 16\% \\
& AutoCodeRover & (Claude 3.5 Sonnet, top-3 files) & 43\% \\
& OpenHands & (Claude 4 Sonnet, top-3 files) & 64\% \\
& OrcaLoca & (Claude 4 Sonnet, top-3 files) & 45\% \\
& Agentless & (Gemini 2.5 Pro, top-3 files) & 36\% \\
& RGFL & (Gemini 2.5 Pro, top-3 files) & \textbf{69\%} \\
\midrule
\multirow{5}{*}{Lite} 
& RAG & (BM25) & 15\% \\
& AutoCodeRover & (GPT-4o, top-3 files) & 32\% \\
& OpenHands & (Claude 3.5 Sonnet, top-3 files) & \textbf{71\%} \\
& OrcaLoca & (Claude 4 Sonnet, top-3 files) & 43\% \\
& Agentless & (Gemini 2.5 Pro, top-3 files) & 20\% \\
& RGFL & (Gemini 2.5 Pro, top-3 files) & 68\% \\
\midrule
\multirow{4}{*}{Java} 
& RAG & (BM25) & 14\% \\
& OpenHands & (Claude 3.7 Sonnet, top-3 files) & 38\% \\
& OrcaLoca & (Claude 4 Sonnet, top-3 files) & 9\% \\
& Agentless & (Gemini 2.5 Pro, top-3 files) & 31\% \\
& RGFL & (Gemini 2.5 Pro, top-3 files) & \textbf{46\%} \\
\bottomrule
\end{tabular}
\end{adjustbox}
\end{table*}

\mybox{\textbf{Answer to RQ2:} Our reasoning-guided fault localization approach generalizes effectively across datasets and programming languages, while outperforming or matching strong baselines. At the file level, our method consistently achieves the best Hit@3, Recall@3, and MRR across SWE-bench Verified, Lite, and Java. Although OrcaLoca sometimes has an edge at Hit@1, our approach provides stronger overall rankings, which is more impactful for downstream repair. At the element level, our method achieves the highest exact match on Verified and Java, and remains competitive on Lite.}

\subsection{RQ3: Does ranking files and elements based on LLM reasoning improve the success rate of final program repairs compared to baseline agents?}

For this research question, we first ran the original Agentless system with Gemini 2.5 Pro across all 500 instances of the SWE-bench Verified dataset to establish a baseline. We then re-ran the repair pipeline after substituting in our reasoning-based file localization, and again after adding our reasoning-based element localization. The results are summarized in Table~\ref{tab:resolved_instances}.

For a broader context, we also include results from other agents on the SWE-bench Verified leaderboard. OpenHands with Claude 4 Sonnet currently reports the strongest results. However, since model choice can heavily influence outcomes, we also report OpenHands performance using Gemini 2.5 Pro for a more direct comparison.

The results reveal three key findings:
\begin{itemize}
    \item Agentless vs. OpenHands: When using Gemini 2.5 Pro as the LLM backend, Agentless outperforms OpenHands, resolving 258/500 instances (52\%) compared to OpenHands’ 49\%, which may indicate that OpenHands's improvement is very much LLM-dependent (in this case, fine-tuned for Claude)
    \item Impact of reasoning-based file localization: Substituting our file localization increases the number of resolved issues from 258 to 279, demonstrating that more accurate file selection leads to a measurable increase in repair success.
    \item Impact of reasoning-based file + element localization: Incorporating both our file and element localization further boosts performance to 291/500, showing that improvements at both levels compound to meaningfully enhance end-to-end repair outcomes.
\end{itemize}

\begin{table}[t]
\centering
\caption{Percentage of resolved instances across different methods on SWE-bench Verified}
\label{tab:resolved_instances}
\begin{adjustbox}{max width=0.85\linewidth}
\begin{tabular}{@{}lr@{}}
\rowcolor{gray!20}
\textbf{Method} & \textbf{Resolved (\%)} \\
\toprule
AutoCodeRover (Claude 3.5 Sonnet) & 46.2\\
OpenHands (Claude 4 Sonnet) & 70.4\\
OpenHands (Gemini 2.5 Pro) & 49 \\
Agentless (Gemini 2.5 Pro) & 51.6 \\
RGFL file loc + Agentless element loc (Gemini 2.5 Pro) & 55.8 \\
RGFL file loc + RGFL element loc & 58.2  \\
\bottomrule
\end{tabular}
\end{adjustbox}
\end{table}

\mybox{\textbf{Answer to RQ3:} LLM reasoning in file and element localization directly improves the final success rate of program repair. On SWE-bench Verified, adding reasoning-based file localization increased resolved cases from 258 (51.6\%) to 279 (55.8\%), and combining reasoning for both files and elements further increases this to 291 (58.2\%). This shows that better localization meaningfully compounds into higher end-to-end repair rates.}

\subsection{Discussion: Counterfactual ablation}
\begin{figure*}
    \centering
    \includegraphics[width=0.7\linewidth]{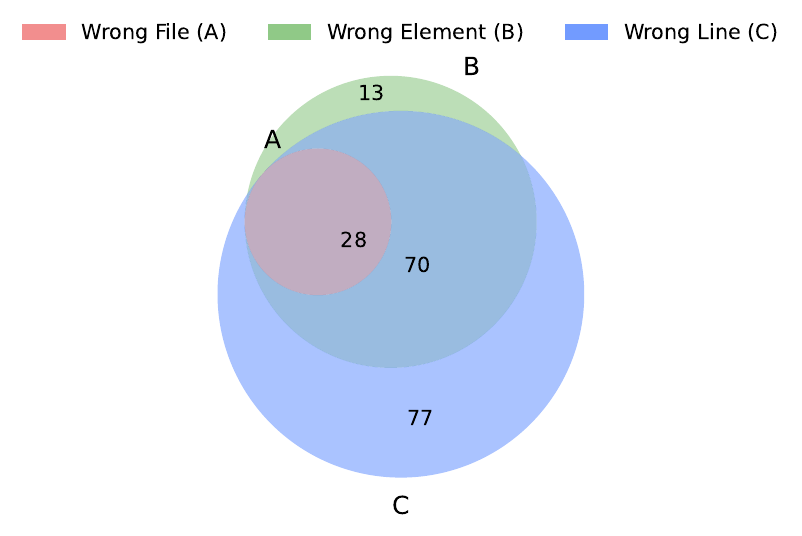}
    \caption{The distribution of unresolved cases across localization stages. (file, element, and line level). There are in total 28 issues with wrong localized files, 111 (28+13+70) issues with wrong localized elements, and 175 (28+70+77) issues that have wrong localized lines, in our generated patches}
    \label{fig:fig4}
\end{figure*}

To understand the sources of error in our pipeline, we conducted a counterfactual ablation on the 209 unresolved SWE-bench Verified instances. In Figure~\ref{fig:fig4}, you can see the Venn diagram\cite{venn1880diagrammatic} of this analysis. Error attribution (not mutually exclusive) shows: wrong file in 28 cases (13\%), wrong element in 111 cases (53\%), and wrong line in 175 cases (84\%). We also observed a subset of 21 instances where file/element/line localization were all correct, but the final patch still failed—indicating pure patch-generation/validation errors.

We then asked: how much could accuracy improve if each localization stage were hypothetically corrected with ground-truth (GT) information? To test this, for each localization phase, we replaced the wrong localized file/element/line with the GT step-by-step and re-ran the downstream pipeline unchanged. We observed the following changes per ablation step:
\begin{itemize}
    \item GT files: Replacing the retrieved files with ground-truth ones led to 14 of the 28 file-miss cases being resolved (50\%). This means that although wrong file localization is less likely, given our approach, filling the gap to the perfect localization yields the highest payoff per instance.
    
    \item GT files + GT elements: In the 111 cases where element localization was wrong, replacing the retrieved elements with ground-truth elements (while keeping the GT files) led to 29 cases being resolved ($\approx$26\%). This shows that element accuracy contributes to repair success, but downstream line localization and patch generation remain significant bottlenecks.
    
    \item GT files + GT elements + GT lines: In the 175 cases where line localization was wrong, replacing the retrieved lines with ground-truth lines (while keeping earlier stages fixed) led to 33 cases being resolved ($\approx$19\%). This shows that line errors are the most frequent overall, but correcting them alone yields the smallest recovery, suggesting that many unresolved cases stem from limitations in patch generation rather than localization.
\end{itemize}

Overall, this ablation highlights that correcting file localization has the largest per-case impact, while line localization errors are the most common but hardest to fix in isolation.

Building on this analysis, we also examined whether element-level localization is always necessary, once the correct file has been identified. Specifically, we conducted an ablation study on the unresolved cases where file localization was correct but element localization was wrong, and even after injecting the ground-truth elements, these cases still failed to resolve (which are 70 samples). We then tested whether providing only the ground-truth files—without further element or line localization—could help. Interestingly, in this setting, 9 cases were repaired. This suggests that for some instances, giving the model a precise element to edit may actually reduce the model's repair power, while simply constraining the repair process to the correct file can be sufficient for resolving the bug. 

For instance, there is a sample in SWE-bench Verified where "django/db/models/expressions.py" is the ground-truth file, and the fix adds a new method: get\_group\_by\_cols(self, alias=None). When we provided the file, the model followed the surrounding patterns and generated the correct patch with the alias parameter. However, when we emphasized only the ground-truth elements, the model over-focused on these and ignored broader patterns in the file, generating a wrong version of the method without the alias parameter. This shows that narrowing the context to specific elements can sometimes confuse the model, while broader file-level guidance can lead to the correct repair.

Therefore, we conclude that finer-grained localization (e.g., at the element and line levels) deserves further attention from the research community—not only because it is a more challenging problem than file-level localization, but also because the optimal way of presenting this information in the prompt remains unclear. Striking the right balance is crucial to guide the model without overly limit its exploration span.

We also plan an empirical study to quantify the impact of file, element, and line localization on final repair performance, and to analyze how and to what extent additional contextual information improves repair outcomes.

\begin{comment}
\begin{figure*}[t]
    \centering
    \begin{subfigure}{0.48\linewidth}
        \centering
        \includegraphics[width=\linewidth]{figures/venn_unresolved.pdf}
        \caption{}
        \label{fig:venn_unresolved}
    \end{subfigure}
    \hfill
    \begin{subfigure}{0.48\linewidth}
        \centering
        \includegraphics[width=\linewidth]{figures/venn_resolved_with_gt_compact.pdf}
        \caption{}
        \label{fig:venn_resolved}
    \end{subfigure}
    \caption{Error attribution and counterfactual resolution analysis. 
    (a) shows how unresolved cases are distributed across localization stages. (file, element, and line level)
    (b) shows how many of those cases would be resolved if ground truth information were injected at each stage. (file, element, and line level)}
    \label{fig:fig4}
\end{figure*}
\end{comment}

\section{Threats to Validity}
Our experiments focus on Python (SWE-bench Verified and Lite) and Java (SWE-bench Java). While this covers multiple languages, the results may not fully generalize to other ecosystems such as C++, JavaScript/TypeScript, or Rust. Future work should include additional languages and repositories to better evaluate cross-language applicability.

We selected Gemini 2.5 Pro as the LLM backend in later stages because it offered a favorable balance of accuracy and cost during our experiments. However, this choice may not hold universally—other LLMs or future model releases could perform differently. For baselines, we relied on the best-reported LLMs at the time, which may not correspond to the current best-performing models.

For patch validation, we rely solely on the regression tests included in the original repositories. While this is standard practice in SWE-bench, it means we do not use additional reproduction tests.

For RQ3, we report repair performance only on SWE-bench Verified. This provides a reliable benchmark but limits the scope of our conclusions. Extending repair evaluation to SWE-bench Lite and SWE-bench Java would strengthen the generality of our findings and provide more complete evidence of how RGFL impacts downstream repair across datasets.
\section{Conclusion and Future Work}
In this paper, we propose a novel LLM-based reasoning framework for fault localization(RGFL), aimed at improving both file and element-level localization in APR. Unlike traditional spectrum-based or retrieval-based methods, our approach uses LLMs not just for code generation, but as reasoning engines capable of interpreting bug reports and explaining code functionality in natural language. These reasonings are then used to guide the localization process more accurately.

Through extensive evaluation on the SWE-bench dataset, we showed that incorporating LLM-based reasoning both at the file level and element level localization leads to consistent improvements over strong baselines. In addition, these improvements in fault localization translate into significant gains in end-to-end repair success, as shown by our experiments where RGFL improved the number of bugs fully resolved. %These results validate that LLM-generated explanations offer meaningful semantic signals that can be used to reason about bug-report-to-code relevance, surpassing models that rely only on similarity or structural signals.

While this work focuses on improving fault localization within APR, the broader idea of LLM-based reasoning can be extended to other areas of software engineering. Potential applications include test case generation, code summarization, specification mining, code review assistance, and vulnerability detection. Each of these tasks requires understanding both code and intent, a setting where LLM-generated reasoning can play a powerful role.
\section{Data Availability}

We release the source code of our experiments to help other researchers replicate and extend our study.\url{https://github.com/MelikaSepidband/RGFL}

\balance
%\bibliographystyle{IEEEtran}
\bibliographystyle{ACM-Reference-Format}

\bibliography{refs}
\end{document}